\documentstyle[11pt]{article}
\newcommand{\bdis}{\begin{displaymath}}
\newcommand{\edis}{\end{displaymath}}
\newcommand{\beqn}{\begin{equation}}
\newcommand{\eeqn}{\end{equation}}
\newcommand{\beqna}{\begin{eqnarray}}
\newcommand{\eeqna}{\end{eqnarray}}
\newcommand{\nn}{\nonumber}

\newcommand{\epc}{\epsilon}

\newcommand{\del}{\partial}
\newcommand{\dcsb}{D$\chi$SB}
\def\NP(#1,#2,#3){Nucl. Phys. {\bf #1}(19#2)#3}
\def\NPBPS(#1,#2,#3){Nucl. Phys. {\bf B}(Proc. Suppl.){\bf #1}(19#2)#3}
\def\PL(#1,#2,#3){Phys. Lett. {\bf #1}(19#2)#3}
\def\PR(#1,#2,#3){Phys. Rev. {\bf #1}(19#2)#3}
\def\PRL(#1,#2,#3){Phys. Rev. Lett. {\bf #1}(19#2)#3}
\def\PTP(#1,#2,#3){Prog. Theor. Phys. {\bf #1}(19#2)#3}
\def\PTPS(#1,#2,#3){Prog. Theor. Phys. (Suppl.){\bf #1}(19#2)#3}

\newcommand{\DGL}{{\rm DGL}}
\newcommand{\QCD}{{\rm QCD}}
\newcommand{\eff}{{\rm eff}}
\newcommand{\conf}{{\rm conf}}
\newcommand{\qk}{{\rm quark}}
\newcommand{\qg}{{\mbox{\scriptsize\rm q-g}}}
\newcommand{\Yu}{{\rm Y}}
\newcommand{\Cou}{{\rm C}}
\newcommand{\lag}{{\cal L}}
\newcommand{\tr}{{\rm tr}}
\newcommand{\Tr}{{\rm Tr}}
\newcommand{\Del}{{\cal D}}
\newcommand{\gpar}{\alpha_{\rm e}}
\newcommand{\Nf}{N_{\rm f}}
\newcommand{\Nc}{N_{\rm c}}
\begin{document}
\title{\bf 
Monopole Dominance for Dynamical Chiral-Symmetry Breaking 
in the Dual Ginzburg-Landau Theory
} 
\author{\normalsize
S.~Umisedo\thanks{e-mail address : umisedo@rcnp.osaka-u.ac.jp},
S.~Sasaki, H.~Suganuma and H.~Toki \\
\\
{\small \it 
Research Center for Nuclear Physics (RCNP), Osaka University, 
Ibaraki, Osaka 567, Japan
}\\
}
\date{}
\maketitle
\begin{abstract}
Using the effective potential formalism,
we study dynamical chiral symmetry breaking
(\dcsb) in the dual Ginzburg-Landau (DGL) theory,
where the color confinement is brought by monopole condensation.
The effective potential as a function of infrared quark mass
is found to have double-well structure, which leads to spontaneous
breaking of chiral symmetry.
To examine the role of confinement,
we divide the effective potential into the confinement part and others,
which correspond to the confinement term and other (Yukawa, Coulomb) terms
of the gluon propagator in the DGL theory.
It is found that the confinement part gives the dominating contribution
to the \dcsb, which is regarded as monopole dominance for \dcsb.
\end{abstract}
\newpage

Quark confinement and dynamical chiral-symmetry breaking(\dcsb) are the
most important nonperturbative phenomena of QCD.
However, the relation between these two phenomena is not yet clear
	\cite{kogut,banks}.
As for confinement, from recent studies of the lattice gauge theory
	[3--5]
, 
it is getting clear that quark confinement can be understood
in terms of the dual superconductor picture in the 't Hooft abelian gauge
	\cite{thooft1}.
In this picture, the QCD-monopole appearing by the abelian gauge fixing
condenses in the QCD vacuum,
and therefore quarks are confined through the dual Meissner effect
	[7--9]
.
The picture also provides a new aspect for understanding \dcsb.
In the effective theory approach for these phenomena
with the dual Ginzburg-Landau(DGL) theory
	\cite{suzuki}, 
which models the dual Meissner effect for confinement,
it was shown that QCD-monopole condensation plays an essential role on \dcsb
	[11--14]
.
At the same time, 
	Miyamura\cite{miyamura} and Woloshyn\cite{woloshyn}
found the similar conclusion in the lattice gauge theory.

The DGL theory is an infrared effective theory of the nonperturbative QCD,
where the QCD vacuum is described in terms of 
the abelian gauge theory with QCD-monopoles
	\cite{suzuki}.
In this theory,
it was shown that monopole condensation
induces large dynamical quark mass using the Schwinger-Dyson(SD) equation,
a non-linear integral equation for the quark mass function
	[11--13]
.
The chiral phase transition at finite temperature was also studied
by using the SD equation.
A strong correlation was found between the critical temperature
of the chiral symmetry restoration and the QCD-monopole condensate
	\cite{sasakiB}.
Since monopole condensation directly leads to confinement,
the above results suggest strong correlation between
confinement and \dcsb.
To get deeper insight on this correlation 
it is desirable to examine the contribution
of confinement to \dcsb ~apart from other contributions.
In the DGL theory, the nonperturbative gluon propagator is composed of
two parts, confinement term and others.
As for the inter quark potential,
the former leads to the linear potential and latter to the Yukawa potential.
To examine the contribution of each part, we calculate the effective potential
at the one-loop level, where
we can simply extract the contribution of confinement on \dcsb.

The dual Ginzburg-Landau (DGL) theory is an infrared effective theory of QCD
based on the dual Higgs mechanism in the abelian gauge
	[10--13]
.
Its Lagrangian is described by quarks $q$, monopoles $\chi$, and diagonal
part of gauge fields as
\beqn
    \lag_\DGL 
	=
    \tr K_{\rm gauge}(A_\mu ,B_\mu )
  + \bar q (i\not\! \del - e \not\!\! A - m) q 
  + \tr [\Del_\mu , \chi ]^\dagger [\Del^\mu , \chi ]
  - \lambda \tr (\chi^\dagger \chi - v^2)^2, 
\label{eqn:DGLlag}
\eeqn
where 
the diagonal gauge field
$A_\mu $ 
and the dual gauge field 
$B_\mu $ 
are defined on the Cartan sub-algebra 
$\vec H=(T_3,T_8)$: 
$A^\mu  \equiv A^\mu _3 T_3 + A^\mu _8 T_8$, 
$B^\mu  \equiv B^\mu _3 T_3 + B^\mu _8 T_8$,
and 
$ \Del_\mu  \equiv  \del_\mu + igB_\mu $ 
is the dual covariant derivative with 
the dual gauge coupling 
$g$
obeying the Dirac condition,
$eg=4\pi $ 
	\cite{suganumaA}.
The QCD-monopole field 
$\chi $ 
is defined on the nontrivial root vectors 
$E_\alpha $:
$\chi  \equiv \sqrt{2} \sum_{\alpha =1}^3 \chi _\alpha E_\alpha $. 
The kinetic term of gauge fields 
$ K_{\rm gauge}(A_\mu ,B_\mu )$ 
is expressed in the Zwanziger form
	\cite{suzuki,suganumaA},
\beqn
 K_{\rm gauge} (A_\mu ,B_\mu ) 
	\equiv 
  - 
    \frac{2}{n^2}
    [n\cdot (\del \wedge A)]^\nu 
    [n\cdot ^*(\del \wedge B)]_\nu 
  - 
    \frac{1}{n^2}
    [n\cdot (\del \wedge A)]^2
  - 
    \frac{1}{n^2}
    [n\cdot (\del \wedge B)]^2, 
\label{eqn:KinGau}
\eeqn
which manifests the duality of the gauge theory. 
The QCD-monopoles condense in the vacuum,due to its self interaction .
QCD-monopole condensation squeezes out the color electric flux 
and leads to linear inter-quark potential
	\cite{suzuki,suganumaA,sasaki}.

In the QCD-monopole condensed vacuum, 
the nonperturbative gluon propagator 
	\cite{suganumaA} 
is derived by integrating out 
$B_\mu $ 
	\cite{suganumaA} 
at the mean field level, 
\beqn
    D_{\mu \nu }(k)
	=
    {1 \over k^2}
	\{
	  g_{\mu \nu }
       	+ (\gpar - 1){k_\mu k_\nu \over k^2} 
       	\}
  + {1 \over k^2} 
	{m_B^2 \over k^2 - m_B^2}
	{1 \over (n\cdot k)^2}
	\epc^\lambda  \, _{\mu \alpha \beta}
	\epc_{\lambda \nu \gamma \delta}
	n^\alpha n^\gamma k^\beta k^\delta , 
\eeqn
where 
$B_\mu $ 
acquires a mass of 
$m_B=\sqrt{3}gv$.
For the study of chiral symmetry, which is a symmetry of light quarks,
we have to take into account the screening effect 
by the light quark polarization
	\cite{suganumaA}.
In this case, the nonperturbative part is modified with the
infrared cutoff parameter 
$a$ 
corresponding to the inverse of hadron size as
	\cite{sasaki} 
\beqn
    D_{\mu \nu }(k)
	=
    {1 \over k^2}
	\{
	  g_{\mu \nu }
       	+ (\gpar - 1){k_\mu k_\nu \over k^2} 
       	\}
  + {1 \over k^2} 
	{m_B^2 \over k^2 - m_B^2}
	{1 \over (n\cdot k)^2 + a^2}
	\epc^\lambda  \, _{\mu \alpha \beta}
	\epc_{\lambda \nu \gamma \delta}
	n^\alpha n^\gamma k^\beta k^\delta .
\label{eqn:DDGL}
\eeqn

It was shown that the role of QCD-monopole condensation 
is essential on D$\chi $SB 
by solving the Schwinger-Dyson (SD) equation 
	[11--13]
. 
Approximating the full quark propagator as 
$iS^{-1}= \not\!\!\, p - M(p^2) + i\epc $, 
one obtains the SD equation for the quark mass function 
$M(p^2)$
in the Euclidean space,
\beqn
    M(p^2)
	= 
    \int\!\!{d^4q \over (2\pi )^4} 
    \vec Q^2 {M(q^2) \over q^2 + M^2(q^2)} 
    D_{\mu \mu}(q-p)
\label{eqn:SDE}
\eeqn
in the chiral limit.
Here,
$\vec{Q}^2
	=
	{\Nc - 1  \over  2\Nc}e^2
$
denotes the abelian electric charge of the quark.
In this equation,
$D_{\mu \mu }(k)$ 
is composed of three parts, 
\beqn
    D_{\mu \mu}(k)
	= 
   {2m_B^2 \over k^2(k^2 + m_B^2)} \cdot
   {n^2k^2 + a^2 \over (n \cdot k)^2 + a^2}
  +{2 \over k^2 + m_B^2}
  +{1 + \gpar \over k^2}
	=
    D_{\mu \mu}^\conf(k)
  + D_{\mu \mu}^\Yu(k)
  + D_{\mu \mu}^\Cou(k).
\label{eqn:GpropTr}
\eeqn
The first term 
$D_{\mu \mu }^\conf(k)$ 
is responsible for the linear confinement potential 
	\cite{como,suganumaA} 
in the absence of the screening effect,
$a=0$. 
The second term 
$D_{\mu \mu }^\Yu(k)$ 
is related to the short-range Yukawa potential,
and the Coulomb part 
$D_{\mu \mu }^\Cou(k)$ 
does not contribute to the quark static potential. 
In spite of this decomposition,
it is difficult to separate and compare each contribution 
to \dcsb~in the nonlinear SD equation.
To examine these contributions separately,
it is useful to study the effective potential
which corresponds to the vacuum energy density.
In this paper, we study the effective potential
	\cite{cjt}
variationally in the chiral limit.

Within the two-loop diagram approximation, 
the effective potential 
$V_\eff[S]$ 
leads to the ladder SD equation when imposed the extremum condition 
in terms of the full quark propagator 
$S(p)$ 
	\cite{higashijima}.
Using the nonperturbative gluon propagator 
$D^{\mu \nu}(k)$ 
in the DGL theory, the effective potential as a functional of
the dynamical quark mass 
$M(p^2)$ 
is expressed as 
\beqn
    V_\eff[S]
    	=
    i\Tr \ln[S_0 S^{-1}]
  + i\Tr [SS_0^{-1}]
  + \int\!\! {d^4p \over (2\pi)^4}{d^4q \over (2\pi)^4}
   {\vec Q^2 \over 2}
    \tr[\gamma _\mu  S(p) \gamma _\nu  S(q) D^{\mu \nu}(p-q) ],
\label{eqn:EffAct}
\eeqn
where 
$S_0 (p)$ 
is the bare quark propagator, 
$iS_0^{-1}(p) = \not\!\!\, p - m + i\epc $ 
. 

Here, we would like to comment on neglecting the wavefunction renormalization.
Considering the effect of the wavefunction renormalization, the ladder
SD equation for the quark propagator,
$iS^{-1}(p)=z^{-1}(p^2)(\not\! p - M(p^2) ) $,
is given by coupled equations
\beqna
	M(p^2)
	  & = &
     	z(p^2)
     	\int\!\! {d^4q \over (2\pi)^4i} \vec{Q}^2 z(q^2)
     	\frac{M(q^2) g^{\mu \nu}}
     	     {q^2 - M^2(q^2)}
     	D_{\mu \nu},		\label{eqn:sigma}	\\
     	z^{-1}(p^2)
     	  & = &
     	1
     -	{1 \over p^2}
     	\int\!\! {d^4q \over (2\pi)^4i} \vec{Q}^2 z(q^2)
     	\frac{p^\mu q^\nu + p^\nu q^\mu - p \!\cdot\! q\,g^{\mu \nu}}
     	     {q^2 - M^2(q^2)}
     	D_{\mu \nu}.		\label{eqn:z}
\eeqna
Suppose one uses the perturbative gluon propagator and 
the Higashijima--Miransky approximation 
	\cite{higashijima,miransky} 
for the running coupling constant
\beqn
	e^2(k^2)
	\simeq
	e^2(\bar{k}^2),
\eeqn
where 
$k=p-q$, $\bar{k}^2 \equiv \max(p^2,q^2)$.
Then, Eq.(\ref{eqn:z}) reduces to the form
\beqn
	z^{-1}(p^2)
	  =
	1
     +  \gpar{2 \over p^2}
     	\int\!\! {d^4q \over (2\pi)^4i} \vec{Q}^2(\bar{k}^2) 
     	\frac{z(q^2)}
     	     {q^2 - M^2(q^2)}
     	{p\cdot q  \over (p-q)^2},
\eeqn
which leads
$z(p^2) \equiv 1$
in the Landau gauge
($\gpar = 0$).
In this case, Eq.(\ref{eqn:sigma}) reduces to Eq.(\ref{eqn:SDE}).
On the other hand, with the use of the nonperturbative gluon propagator in 
Eq.(\ref{eqn:DDGL}), this is not necessarily the case.
However, we are not interested in the numerical details here,
but in the essential contributions alone.
Hence, working in the Landau gauge,
we neglect the effect of the wavefunction renormalization.
In other words, we neglect the second term in the right hand side of  
Eq.(\ref{eqn:z}) and set
$z(p^2) \equiv 1$ there.

In the effective potential formalism, its 2-loop term is given as
\beqna
	V^{(2)}
	&=& 
	2\Nf\Nc\!\!
	\int\!\!\!{d^4q \over (2\pi)^4i}\!\!
	\int\!\!\!{d^4p \over (2\pi)^4i}
	\vec{Q}^2 z(q^2) z(p^2)
	\frac{M(q^2)M(p^2)g^{\mu \nu}
	    + p^\mu q^\nu  + p^\nu q^\mu
	    - p \!\cdot\! q \,g^{\mu \nu}
	      }
	     {(M^2(q^2) - q^2)
	      (M^2(p^2) - p^2)
	      }
	D_{\mu \nu}					\nn\\		
	&=&
	2\Nf\Nc\!\!
	\int\!\!\!{d^4q \over (2\pi)^4i}\!\!
	\int\!\!\!{d^4p \over (2\pi)^4i}
	\vec{Q}^2 z(q^2) z(p^2)
	\frac{M(q^2)M(p^2)}
	     {(M^2(q^2) - q^2)(M^2(p^2) - p^2)}
	D^\mu{}_{\mu}					\nn\\
	&+&
	2\Nf\Nc\!\!
	\int\!\!\!{d^4p \over (2\pi)^4i}
	\frac{p^2}
	     {M^2(p^2) - p^2}
	\{1-z(p^2)\},
\label{eqn:V2true}
\eeqna
where Eq.(\ref{eqn:z}) is used.
Here, to neglect the effect of the wavefunction renormalization and
set $z(p^2) \equiv 1$ is equivalent to
neglect other terms than that contains
$M$ in the numerator of Eq.(\ref{eqn:V2true}).
We expect that this approximation holds good in the Landau gauge
as it exactly does in the QCD-like case
	\cite{higashijima,miransky}.
Hence, the effective potential consistent with the approximation
$z(p^2) \equiv 1$ is given by 
\beqna
    V_\eff[M(p^2)]
	&\equiv&
    V_\qk[M(p^2)] + V_\qg[M(p^2)]		\nn \\
    	&=&
  - 2\Nf \Nc 
    \int\!\! {d^4p \over (2\pi)^4}
    \{
      \ln({p^2 + M^2(p^2) \over p^2})
     -2{M^2(p^2) \over p^2 + M^2(p^2)}
    \}						\nn \\
	&-&
    \Nf(\Nc - 1) 
    \int\!\! {d^4p \over (2\pi)^4} {d^4q \over (2\pi)^4}
    e^2 {M(p^2) \over p^2 + M^2(p^2)} 
   {M(q^2) \over q^2 + M^2(q^2)} 
    D_{\mu \mu}(p-q)  \label{eqn:EffPot}  	  
\eeqna
in the Euclidean space.
It is easy to check that the extremum condition on Eq.(\ref{eqn:EffPot})
with respect to $M(p^2)$ leads to the SD equation (\ref{eqn:SDE})
	\cite{suganumaA,sasaki}.

In Eq.(\ref{eqn:EffPot}),
the first term $V_\qk$ is the one-loop contribution as in Fig.1a.
The second term $V_\qg$ with 
$D_{\mu \mu }$ 
is the two-loop contribution with the quark-gluon interaction
as expressed in Fig.1b.
The important point is that this second term
$V_\qg$
is divided into three parts as
\beqn
	V_\qg
	=
	V_\conf + V_\Yu + V_\Cou ,
\eeqn
corresponding to the decomposition of 
$D_{\mu \mu }$ 
in Eq.(\ref{eqn:GpropTr}).
Hence, by estimating
$V_\conf$, $V_\Yu$, $V_\Cou$
respectively, it is possible to examine each contribution to \dcsb.
Before going further, remember that 
the nonperturbative gluon propagator depends on the Dirac string direction
$n_\mu$.
However, due to the
$q$-$\bar{q}$
pair polarization effect around the quark,
the Dirac-string direction 
$n_\mu $
becomes indefinite
	\cite{suganumaA}.
Therefore we take the average value on
$n_\mu $,
\beqn
	\langle
	{n_\mu n_\nu \over
	 (n \cdot k)^2 + a^2 }
	\rangle _{\rm av}
   =
   	{1 \over 2\pi^2}
   	\int \!\!\! d\Omega_n
   	{n_\mu n_\nu \over
	 (n \cdot k)^2 + a^2 }
   =
   	{k_\mu k_\nu \over k^2}f_{/\!\!/}(k^2)
   +	\left\{ \delta_{\mu \nu}
   		- {k_\mu k_\nu \over k^2}
   	\right\}f_{\perp}(k^2),
\eeqn
with
\beqna
	f_{/\!\!/}(k^2)
  &=&
   	{1 \over (a + \sqrt{a^2 + k^2} )^2 },		\\
	f_{\perp}(k^2)
  &=&
  	\frac{ a + 2\sqrt{a^2 + k^2} }
  	     { 3a (a + \sqrt{a^2 + k^2} )^2 }
   =	{1 \over 3}
   	\left\{
   		{2(a^2+k^2) \over ak^2(a + \sqrt{a^2 + k^2})}
   		-
   		{1 \over k^2} 
   	\right\},
\eeqna   	
where the angle integration is performed in the Euclidean
4-dimensional space.
We then obtain the averaged gluon propagator 
$\bar{D}_{\mu \nu}$ as
\beqn
	\bar{D}_{\mu \nu}
   =	
   	{1 \over 3}
   	\left\{	\delta_{\mu \nu}
   		-
   		{k_\mu k_\nu \over k^2}
   	\right\}
   	\left\{	{1 \over k^2}
   		+
   		{2 \over k^2 + m_B^2}
   		+
   		{4m_B^2 \over k^2(k^2 + m_B^2)}
   		{a^2 + k^2 \over a(a + \sqrt{a^2 + k^2})}
   	\right\}
   +	
   	\gpar{k_\mu k_\nu \over (k^2)^2}
\eeqn
which requires the modification to Eq.(\ref{eqn:GpropTr})
only on confinement part as
\beqn
	D^\conf_{\mu \mu}
	\rightarrow
	\frac{4m_B^2}
	     {k^2 (k^2 + m_B^2)}
	\frac{k^2 + a^2}
	     { a (a + \sqrt{k^2 + a^2} ) }.
\eeqn
Therefore the effective potential we ought to compute is given as the sum of
following four terms
\beqna
	V_\qk \;\,
 & = &  \hspace{10pt}
 	\frac{2N_{\rm f}N_{\rm c}}{(4\pi)^{2}} \!\!\!
        \int^\infty_0\!\!\!\!\!\!\!\! dp^{2}\! 
        \left\{ 
             - p^{2}\ln \left( 1+\frac{M^2(p^2)}{p^{2}} \right) 
             + \frac{2\,p^{2}M^2(p^2)}{M^2(p^2)+p^{2}} 
        \right\} 						 \\
	V_{\rm C} \;
 & = &- \frac{2N_{\rm f}N_{\rm c}}{(4\pi)^{2}} \!\!\!
        \int^\infty_0\!\!\!\!\!\!\!\! dp^{2}\! 
        \frac{p^{2}M(p^2)}{M^2(p^2)+p^{2}} 
        \!\!\int^\infty_0\!\!\!\!\!\!\!\! dq^{2}\! 
        \frac{q^{2}M(q^2)}{M^2(q^2)+q^{2}} 
        \vec{Q}^2
        \frac{1}{\max(p^2,q^2)} 				 \\
	V_{\rm Y} \:
 & = &- \frac{2N_{\rm f}N_{\rm c}}{(4\pi)^{2}} \!\!\!
        \int^\infty_0\!\!\!\!\!\!\!\! dp^{2}\! 
        \frac{p^{2}M(p^2)}{M^2(p^2)+p^{2}} 
        \!\!\int^\infty_0\!\!\!\!\!\!\!\! dq^{2}\! 
        \frac{q^{2}M(q^2)}{M^2(q^2)+q^{2}} 
        \vec{Q}^2
        \frac{4}{\pi} \!\!\int^\pi_0\!\!\!\!\!\! d\theta \sin^{2}\theta \,
        \frac{1}{(p-q)^{2}+m^{2}_{B}} 				 \\
	V_\conf \;
 & = &- \frac{2N_{\rm f}N_{\rm c}}{(4\pi)^{2}} \!\!\!
        \int^\infty_0\!\!\!\!\!\!\!\! dp^{2}\! 
        \frac{p^{2}M(p^2)}{M^2(p^2)+p^{2}} 
        \!\!\int^\infty_0\!\!\!\!\!\!\!\! dq^{2}\! 
        \frac{q^{2}M(q^2)}{M^2(q^2)+q^{2}} 
        \vec{Q}^2			 			\nn \\
 &   &  \hspace{18pt} \times
        \frac{8}{\pi a} \!\!\int^\pi_0\!\!\!\!\!\! d\theta \sin^{2}\theta \,
        \frac{1}{a+\sqrt{(p-q)^{2}+a^{2}}} 
        \frac{m_{B}^{2}((p-q)^{2}+a^{2})}
             {(p-q)^{2} ((p-q)^{2}+m_{B}^{2})}.			
\eeqna

As for the running coupling, we adopt hybrid type one
	\cite{suganumaA}
in the Higashijima--Miransky approximation,
\beqn
	e^2((p-q)^2)
	=
	{48\pi ^2 (\Nc + 1) \over 11\Nc - 2\Nf}
	\{\ln {p^2_c + \max(p^2,q^2) \over \Lambda^2_\QCD}
	  \}
	  ^{-1},
\eeqn
which connects smoothly the perturbative running coupling of QCD and the
infrared effective coupling of the DGL theory
	\cite{sasaki}.
From the renormalization group analysis of QCD 
	\cite{higashijima}, 
the approximate form of the quark-mass function 
$M(p^2)$ 
is expected as 
\beqn
    M(p^2)
    	= 
    M(0) {p_c^2 \over (p_c^2 + p^2)} 
    \left[
       \ln \frac{p_c^2}{\Lambda^2_\QCD}
       \left/
       \ln \frac{p_c^2 + p^2}{\Lambda^2_\QCD}
       \right.
    \right]^
       {1 - {\Nc^2 - 1 \over 2\Nc} \cdot {9 \over 11\Nc - 2\Nf}}.
\label{eqn:MassAns}
\eeqn
Since the exact solution 
$M_{\rm SD}(p^2)$ 
of the SD equation (\ref{eqn:SDE}) 
	\cite{sasaki} 
is well reproduced by this ansatz (\ref{eqn:MassAns})
with 
$M(0) \simeq 0.4$GeV 
and 
$p_c^2 \simeq 10\Lambda _\QCD^2$,
we use this form 
as a variational function of the effective potential.

We show in Fig.2 the effective potential
$V_\eff$
as a function of the infrared effective quark mass
$M(0)$,
using the mass function (\ref{eqn:MassAns}) with
$p_c^2 = 10\Lambda^2_\QCD$.
We have used the same parameters as in Ref.
	\cite{suganumaA,sasaki},
$\lambda =25$, $v=0.126{\rm GeV}$, $e=5.5$ and $a=85$MeV
so as to reproduce the inter-quark potential and 
the flux-tube radius 
$R \simeq 0.4 {\rm fm}$ 
	\cite{suganumaA}.
It takes a minimum at finite
$M(0) \simeq $0.4GeV,
which means that the nontrivial solution is more stable
than the trivial one in terms of the energetical argument.
Hence, chiral symmetry is spontaneously broken.

We show in Fig.3 
$V_\qk, V_\conf, V_\Yu$ and $V_\Cou$
as the function of $M(0)$.
The lowering of the effective potential contributes to \dcsb.
Although both the Yukawa and the Coulomb terms contribute slightly to
lower the effective potential, it is mainly lowered by the 
confinement part
$V_\conf$
and there is a large cancellation between $V_\qk$ and $V_\conf$.
Thus, \dcsb~is brought by 
$V_\conf$
arising from monopole condensation in the DGL theory
which is regarded as  monopole dominance for \dcsb.
Such a dominant role of the confinement effect on \dcsb~is found
for any value of $M(0)$.

In Fig.4, we show also the integrands of each term
\beqn
	V_\eff
	=
	\int^\infty_0 \!\!\!\!\!\!
	dp^2 v_\eff(p^2)
	=
	\int^\infty_0 \!\!\!\!\!\!
	dp^2 \{v_\qk(p^2)
     +  v_\conf(p^2)
     +  v_\Yu(p^2)
     +  v_\Cou(p^2)\}
\eeqn
to examine which energy region is important to \dcsb.
Here we have used the exact solution 
$M_{\rm SD}(p^2)$
of the SD equation (\ref{eqn:SDE})
	\cite{suganumaA,sasaki} 
as the mass function to get rid of the ambiguity 
arising from the choice of the trial mass function.
Among $v_\conf$, $v_\Yu$ and $v_\Cou$ 
the confinement part $v_\conf$ is always dominant for all momentum region.
All the three terms, $v_\conf,v_\Yu,$ and $v_\Cou$,
contribute to the effective potential mainly in the low momentum region 
less than 1GeV although there are also long tails running into
high momentum region over 1GeV.
These behaviors directly reflect the
profile of the quark mass function $M_{\rm SD}(p^2)$.
It is notable that such contributions from high momentum region are 
strongly canceled in $v_\eff$ as shown in Fig.4.
Consequently, the remaining contribution only from low momentum region
($p^2 < 0.4 {\rm GeV}^2$)
plays an important role to \dcsb.
This seems consistent with the above result that infrared confinement
effect is dominant for \dcsb.

In summary, we have studied \dcsb~in the DGL theory 
using the effective potential formalism.
Within the two-loop diagram approximation, the effective potential is 
formulated as a function of the dynamical quark mass
$M(p^2)$
neglecting the effect of the wavefunction renormalization.
The effective potential has been calculated as a function
of the infrared quark mass 
$M(0)$
with the variational function (\ref{eqn:MassAns}).
We have found the double-well structure of the effective potential,
so that the nontrivial solution is more stable
than the trivial one and leads to \dcsb.
To examine the role of confinement, the interaction term
$V_\qg$
has been divided into the confinement part
$V_\conf$
and others
($V_\Yu$,$V_\Cou$).
We have found that the confinement part
$V_\conf$
stemming from monopole condensation gives the dominant contribution to 
$V_\qg$
($V_\qg \simeq V_\conf$),
which means the monopole dominance for \dcsb.
It has been also found that the low momentum contribution from less than
1 GeV plays an important role for \dcsb.


\newpage
\section*{Figure Captions}
Fig.1: The diagrams which contribute to the effective potential
up to the two-loop level.
	(a) The quark loop contribution $V_\qk$ 
	without the explicit interaction.
	(b) The two-loop diagram $V_\qg$ 
	including the quark-gluon interaction.
Here, the curly line with a black dot denotes the nonperturbative gluon
propagator in the DGL theory.

\noindent
Fig.2: The total effective potential $V_\eff$ as a function of the infrared
quark mass $M(0)$.
The nontrivial minimum appears at $M(0)\sim 0.4$GeV, which indicates
dynamical breaking of chiral symmetry.

\noindent
Fig.3: $V_\qk$, $V_\conf$, $V_\Yu$ and $V_\Cou$ are is shown as a function
of $M(0)$. The confinement part $V_\conf$, 
plays the dominant role through the lowering the effective potential.

\noindent
Fig.4: Integrands $v_\qk$, $v_\conf$, $v_\Yu$ and $v_\Cou$
of effective potential are shown as functions of the Euclidean momentum $p^2$. 
The confinement part $v_\conf$ is more significant than $v_\Yu$ and $v_\Cou$
for all momentum region.

\end{document}